\documentclass[man]{apa7}

\usepackage[american]{babel}
\usepackage{amsmath}
\usepackage{csquotes}
\usepackage{verbatim}
\usepackage{listings}
\usepackage{xcolor}
\usepackage{multirow} 
\usepackage{pdflscape}   
\usepackage{booktabs}    
\usepackage{pbox}
\usepackage{rotating}
\usepackage{geometry}
\usepackage{adjustbox}
\usepackage{pifont}
\usepackage{array}      
\usepackage{multirow}
\usepackage{setspace}
\definecolor{codegreen}{rgb}{0,0.6,0}
\definecolor{codegray}{rgb}{0.5,0.5,0.5}
\definecolor{codebrown}{rgb}{0.58,0,0.82}
\definecolor{backcolour}{rgb}{0.95,0.95,0.92}

\lstdefinestyle{mystyle}{
    backgroundcolor=\color{backcolour},   
    commentstyle=\color{codegreen},
    numberstyle=\tiny\color{codegray},
    stringstyle=\color{codegreen},
    basicstyle=\ttfamily\footnotesize,
    breakatwhitespace=false,         
    breaklines=true,                 
    captionpos=b,                    
    keepspaces=true,                 
    numbers=left,                    
    numbersep=5pt,                  
    showspaces=false,                
    showstringspaces=false,
    showtabs=false,                  
    tabsize=2
}

\usepackage{amssymb}
\usepackage[style=apa,sortcites=true,sorting=nyt,backend=biber]{biblatex}
\DeclareLanguageMapping{american}{american-apa}
\title{Factor- and Composite-Based Structural Equation Modeling - A New Approach to Incorporate Composites in the Traditional SEM Framework}
\shorttitle{Factor- and Composite-Based SEM}

\authorsnames[{1},{2},{2,3},{4}]{Tamara Schamberger, Florian Schuberth, Jörg Henseler, Yves Rosseel}
\authorsaffiliations{
{Bielefeld University, Faculty of Business Administration and Management, Bielefeld, 33615, Germany},
{University of Twente, Department of Design, Production and Management, Enschede, 7500 AE, The Netherlands},
{Universidade Nova de Lisboa, Nova Information Management School, Campus de Campolide, 1070-312 Lisboa, Portugal},
{Ghent University, Department of Data Analysis, 9000 Ghent, Belgium}}

\leftheader{}

\abstract{
Structural equation modeling (SEM) is a prevalent approach for studying constructs.
Traditionally, these constructs are modeled as reflectively measured latent variables -- common factors that account for the variance-covariance structure of their associated indicators. 
Over the past two decades, there has been growing interest in an alternative way of modeling constructs: the composite, i.e., a linear combination of indicators. 
However, existing approaches to estimating composite models either limit researchers from fully leveraging SEM's capabilities, such as handling missing data, evaluating overall model fit, and testing group differences, or significantly increase complexity of the model specification by introducing additional variables.
Against this background, this paper presents a new way of integrating both common factors and composites in the traditional SEM framework.
Our presented model specification, along with its model-implied variance-covariance matrix, enables researchers to: (i) utilize well-established SEM estimators, including maximum likelihood and generalized least squares estimators, and (ii) can leverage developments from the traditional SEM framework in terms of model specification, evaluation, and handling of missing data.
This way of analyzing structural equation models involving common factors and composites is referred to as factor- and composite-based SEM (FC-SEM).
This advancement aims to enhance the flexibility and applicability of SEM in analyzing constructs.
}


\authornote{
   \addORCIDlink{Tamara Schamberger}{0000-0002-7845-784X}

  Correspondence concerning this article should be addressed to Tamara Schamberger, Department Empirical Methods, Bielefeld University, Universitaetsstrasse 25, 33615 Bielefeld, Germany;
  tamara.schamberger@uni-bielefeld.de \\
  Yves Rosseel and Tamara Schamberger acknowledge that this work was supported by the FWO (Fonds voor Wetenschappelijk Onderzoek – Vlaanderen) under the project number G073026N.\\
  Jörg Henseler acknowledges that this work was supported by national funds through FCT (Fundação para a Ciência e a Tecnologia), under the project - UID/04152/2025 - Centro de Investigação em Gestão de Informação (MagIC)/NOVA IMS - https://doi.org/10.54499/UID/04152/2025 <https://doi.org/10.54499/UID/04152/2025>  (2025-01-01/2028-12-31) and UID/PRR/04152/2025 https://doi.org/10.54499/UID/PRR/04152/2025 <https://doi.org/10.54499/UID/PRR/04152/2025>  (2025-01-01/ 2026-06-30).\\
  During the preparation of this study, we used ChatGPT to improve the readability and language of the manuscript. After using this tool, the authors have reviewed and edited the content as needed and take full responsibility for the publication's content.\\
  A preprint of this manuscript is available on arXiv: https://doi.org/10.48550/arXiv.2508.06112.}

\begin{document}
\maketitle

Structural equation modeling (SEM) is a widely used statistical technique to study constructs and their relationships, i.e., relationships between constructs and their relationships with observed variables, so called indicators \parencite{Bollen1989}.
Its versatility has made SEM indispensable in various disciplines, including psychology \parencite{MacCallum2000}, criminology \parencite{Higgins2002}, and business research \parencite{Hulta}.
By encompassing a variety of statistical techniques such as confirmatory factor analysis \parencite{Joereskog1969}, confirmatory composite analysis \parencite{Schuberth2018}, regression analysis \parencite{Graham2008}, and latent growth curve analysis \parencite{McArdle1988}, SEM provides a comprehensive framework for modeling and testing theories.

\begin{figure}
\centering
    \caption{Example of a Common Factor $\eta$ With $P$ Measures $y_1, \dots y_P$}
    \includegraphics[scale = 0.75]{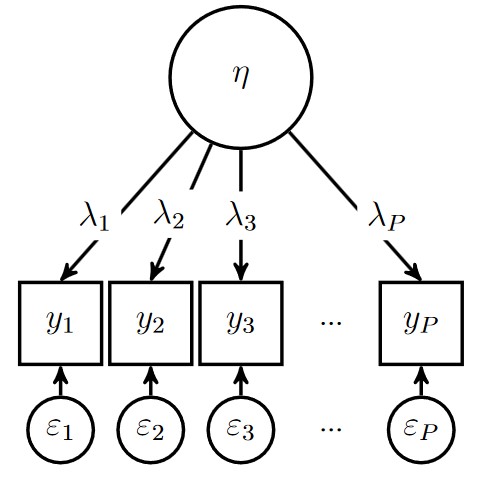}
    \label{fig:LV}
\end{figure}

In research across various disciplines, accurately modeling constructs -- whether constructs that are measured like mental health \parencite[e.g.,][]{Gerino2017} or constructs that are formed like socio-economic status \parencite{Chen2025} -- is crucial \parencite[e.g.,][]{Bollen2011, Strauss2009}.
These constructs, often not directly observed, are inferred from a set of other variables.
Traditionally, researchers model constructs as common factors in SEM, i.e., latent variables explaining the variance-covariance structure of their indicators \parencite{Spearman1904, Joereskog1969}. 
Consequently, common factors are assumed to be the underlying common cause of their indicators.
An example is illustrated in Figure \ref{fig:LV}, where the common factor $\eta$ underlies its measures $y_1, \dots, y_P$.
The variance in a measure $y_i$ that cannot be explained by the common factor is captured by the corresponding random measurement error $\varepsilon_i$ \parencite{Bollen1989}.\footnote{Instead of speaking from random measurement errors, the literature sometimes refers to $\varepsilon_i$ as \textit{unique} factor \parencite[e.g.,][]{Joereskog1969}.}
This approach is particularly useful for modeling constructs that are measured such as intelligence, emotions, or attitudes \parencite{Henseler2021}.

Several approaches to study structural equation models involving common factors have been developed.
Traditionally, estimators that minimize the discrepancy between the model-implied and the sample variance-covariance matrix of the observed variables are used to estimate the model parameters.
Estimating and analyzing structural equation models involving common factors with such approaches is considered the standard across many disciplines.
Therefore, we refer to this type of SEM as the traditional SEM framework in the following sections.
These include approaches such as normal-theory-based generalized least squares \parencite{Joereskog1972, Browne1984}, weighted least squares \parencite{Browne1974} and the maximum likelihood (ML) estimators \parencite{Joereskog1969, Joereskog1970}.
The latter is nowadays regarded as the gold standard for estimating structural equation models \parencite{Hoyle2023a}.
Additionally, alternative approaches have emerged to estimate structural equation models involving common factors.
These approaches rely on composites, i.e., linear combinations of observed variables for this purpose.
This includes methods such as sum score regression \parencite{Richardson1936} and partial least squares path modeling \parencite[PLS-PM,][]{Wold1975}.
Initially touted for their smaller sample requirements and milder distribution assumptions, these approaches are known to yield biased parameter estimates when used for structural equation models involving common factors \parencite[e.g.,][]{Schuberth2023,Schuberth2024, Rigdon2016}.

\begin{figure}
\centering
    \caption{Example of a Composite $\eta$ With $P$ Indicators $y_1, \dots y_P$}
    \includegraphics[scale = 0.75]{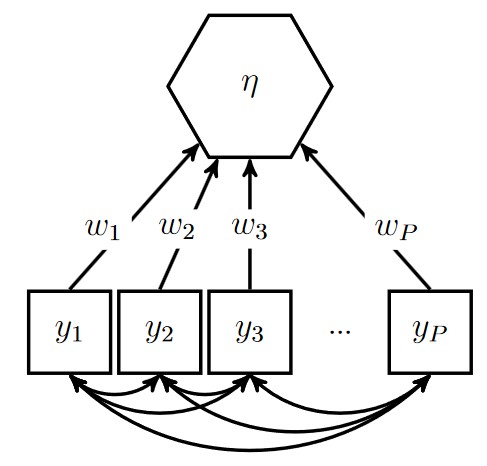}
    \label{fig:composite}
\end{figure}

However, not all constructs are assumed to be the underlying cause of their indicators \parencite{Rhemtulla2020}; for example, constructs such as a company's attractiveness are more plausibly formed by variables such as perceiving the company as successful in attracting high-quality employees, being able to imagine working there oneself and liking the company's physical appearance \parencite{Hair2022}.
When such constructs are nevertheless modeled as common factors, this can result in poor construct validity, where the modeled construct fails to represent the actual construct of interest \parencite{Rhemtulla2020}. 
This misalignment can lead to substantial parameter biases, particularly in relationships between constructs, and adversely affect model assessment \parencite[e.g.,][]{Schuberth2024}.
In such cases, researchers risk drawing wrong conclusions from their analyses.
For these reasons, some constructs are better represented as composites \parencite{Hwang2021a, Schuberth2023e, Cho2022, Henseler2017a} which assume that the construct is composed of its indicators \parencite{Yu2021, Henseler2020}.
Modeling constructs as composites is particularly useful if the construct is formed by humans \parencite{Henseler2021} or if the construct is a collection of heterogeneous causes \parencite{Grace2008}.
For example, extraversion in human development research has been modeled as a composite composed of warmth, gregariousness, assertiveness, activity, excitement seeking, and positive emotions \parencite{Edwards2000}; The firm performance index in information systems research has been modeled as a composite composed of customer relationships, operational excellence, and revenue growth \parencite{Rai2006}; Soil conditions in ecology has been modeled as a composite composed of  soil texture, soil moisture, and soil ph \parencite{Verheyen2003}.
As Figure \ref{fig:composite} illustrates, composites do not impose restrictions on the variance-covariance structure of their indicators, fundamentally distinguishing them from common factors.
Therefore, in line with \textcite{Grace2008}, we illustrate composites as hexagons instead of circles to distinguish them from latent variables.  
Moreover, applications of composites in SEM include item parcels \parencite{Little2002} and fixed-weights composites, i.e., linear combinations of observed variables with given weights, like sum scores \parencite{Schuberth2025}.

As awareness of the different types of constructs has increased, so has the need for approaches to deal with structural equation models involving both composites and common factors.
Estimators that build composites for parameter estimation, such as PLS-PM, generalized structured component analysis \parencite[GSCA,][]{Hwang2004}, and a maximum likelihood (ML) estimator for composite models \parencite[Chapter 6,][]{Schamberger2022} are naturally candidates for this purpose.
However, these approaches are limited to studying structural equation models involving composites.
To overcome these limitations, consistent partial least squares \parencite[PLSc,][]{Dijkstra2015}, integrated generalized structured component analysis \parencite[IGSCA,][]{Hwang2020}, singular value decomposition SEM \parencite[SVD-SEM,][]{Tenenhaus2025}, and a restricted ML estimator \parencite[RML-SEM,][]{Tenenhaus2025} can be used to consistently estimate structural equation models involving common factors and composites. 
However, these approaches do not leverage the full capabilities of approaches developed in the traditional SEM framework, such as model assessment and handling missing data, or can only be used to estimate standardized path coefficients.
Attempts have been made to deal with this issue by incorporating composites in the traditional SEM framework.
These include the two-step approach, the one-step approach \parencite{Grace2008}, the pseudo-indicator approach \parencite{Rose2019}, and the H-O specification \parencite{Henseler2021, Schuberth2023e}.
Yet, these approaches are limited, as they either do not allow composites to be in dependent positions in the structural model (as for the one-step approach), do not allow the composite weights to be freely estimated (as for the pseudo-indicator  and the two-step approach), or they increase the complexity of specifying the model (as for the H-O specification).
To conclude, current approaches to modeling composites and common factors face limitations in terms of model specification, model complexity, model assessment, and the applicability of advanced analytical procedures established in the traditional SEM framework.

To overcome these limitations, we introduce a new way of incorporating composites and common factors jointly in the traditional SEM framework, called \textit{factor- and composite-based SEM} (FC-SEM).
Specifically, we present the model specification and derive the implications of the model on the model-implied variances and covariances of the variables in such a way that (i) estimators developed in the traditional SEM framework can be directly applied, (ii) we can deal with models containing both composites and common factors, and (iii) we allow for different scaling methods for both composites and common factors.
As a result, we can directly rely on enhancements such as model assessment criteria, and approaches to dealing with missing values developed in the traditional SEM framework.
Therefore, our presented approach overcomes the limitations of other approaches to estimate structural equation models involving common factors and composites by allowing for parameter constraints, flexibly modeling composites, and an intuitive model specification.

The remainder of this paper is structured as follows. 
We begin by reviewing the current state of integrating composites and common factors in structural equation models.
We then introduce our approach by explaining the specification, the implications on the variables' model-implied variances and covariances, the identification, the estimation, and the assessment of these models.
Further, our approach is implemented in the open-source R package \texttt{lavaan} \parencite{Rosseel2012, Rosseel2025}.
We discuss this implementation followed by a scenario analysis that illustrates how to specify the model and demonstrates that our approach is able to retrieve the population parameters in the case of a sample that perfectly represents the population.
Moreover, we evaluate the finite sample performance of our approach.
Next, we illustrate our approach by an empirical example that shows that the results of our approach are similar to those of other approaches to study structural equation models with common factors and composites.
Finally, we conclude with a discussion on the advantages and limitations of our approach.

\section{Extant Approaches to Study Common Factors and Composites in a Structural Equation Model}

Traditionally, SEM deals with constructs modeled as common factors.
This type of SEM has a long tradition and remains the default in many disciplines.
Its origins lie in the introduction of factor analysis \parencite{Spearman1904} and path analysis \parencite{Wright1934}. 
Its popularity increased due to the work of Karl G. \textcite{Joereskog1969, Joereskog1970} and the introduction of the software LISREL \parencite{Joereskog1993}. 
Detailed overviews of SEM's history can be found in \textcite{Tarka2017} and \textcite{Bollen2022}.
In contrast, the groundwork to studying constructs as linear combinations of observed variables, i.e., as composites, was laid in the 1980s \parencite[e.g.,][]{Wold1982, Fornell1982}.
Initially, composites were primarily seen as a means for dimension reduction, and researchers did not ascribe any conceptual unity to the components of a composite \parencite{Bollen2011}. 
This perspective shifted with the advent of theoretical frameworks that provided a rationale for modeling constructs using composites \parencite{diamantopoulos2001index, Henseler2017a, henseler2020auxiliary, Sarstedt2016, Fornell1982, Rigdon2012}, alongside the formal establishment of the composite model \parencite{Dijkstra2013c, Dijkstra2017}.

Due to the growing awareness of different ways of modeling constructs, several approaches to deal with structural equation models containing both common factors and composites have been presented.
These can be divided into (i) approaches developed outside the traditional SEM framework and (ii) model specifications aimed at incorporating composites into the traditional SEM framework.
The following two subsections discuss these approaches, including their advantages and disadvantages.
\subsection{Approaches to Analyse Structural Equation Models Involving Common Factors and Composites Developed Outside the Traditional SEM Framework}
The first approach for handling common factors and composites in structural equation models involves using estimators specifically designed for this purpose.
This category includes methods such as PLSc, GSCA, SVD-SEM, and RML-SEM.

PLS-PM is arguably the most widely employed estimator for composite models.
It is a two-step procedure \parencite{Lohmoeller1989, Wold1975}.
In the first step, weights are estimated using the iterative partial least squares algorithm, and are then used to construct composites.
In the second step, relationships between the composites are estimated. 
However, when PLS-PM is applied to models containing common factors, it leads to inconsistent parameter estimates \parencite{Dijkstra1983,Hui1982}.
To address this issue, PLSc was introduced \parencite{Dijkstra2015, Dijkstra2015a}.
PLSc corrects for attenuation bias when common factors are present, offering consistent parameter estimates for models that include both composites and common factors \parencite{Schuberth2018a}.
Despite this advantage, PLS-PM and PLSc are limited in several regards. 
There is little flexibility in specifying models \parencite{Schuberth2023a}.
Specifically, only some error correlations can be freed \parencite{Rademaker2019} and PLS-PM and PLSc do not allow researchers to impose constraints on model parameters \parencite{Schuberth2023a}.
Finally, no closed-form expression for the standard errors have been derived and thus statistical inference for PLS-PM and PLSc relies on bootstrap which increases computational time.

As an alternative to PLS-PM, GSCA has been proposed \parencite{Hwang2014, Hwang2004}. 
GSCA originally estimates models with composites by minimizing the sum of squared residuals using an alternating least squares algorithm. 
Similar to PLS-PM, GSCA produces inconsistent parameter estimates for models involving common factors \parencite{henseler2012generalized}.
To address this issue, GSCA with uniqueness terms for accommodating measurement error was introduced \parencite{Hwang2017}.
In addition, for models containing both common factors and composites, Integrated GSCA \parencite{Hwang2021a} can be used to obtain consistent parameter estimates. 
While these approaches offer flexibility in model specification, they remain limited in terms of model assessment due to their reliance on bootstrap for inference and the limited availability of overall model fit assessment criteria.

Moreover, SVD-SEM has been proposed as an alternative to PLS-PM \parencite{Tenenhaus2025}.
This is a non-iterative approach, that aligns with the framework proposed by \textcite{Dhaene2023} on non-iterative estimators in the structural after measurement \parencite[SAM,][]{Rosseel2024} approach to SEM. 
Starting from a correlation matrix, SVD-SEM yields parameters that are consistent and asymptotically normal through a series of singular value decompositions.
This approach produces standardized parameter estimates only.

Finally, RML-SEM has been introduced \parencite{Tenenhaus2025}.
This approach estimates model parameters by minimizing the fit function for structural equation models involving common factors \parencite{Joereskog1969}, under the constraint that all composites and common factors have a unit variance. 
However, this approach comes with several disadvantages.
First, only standardized estimates can be obtained for the structural parameters.
Second, this approach relies on constrained maximum likelihood estimation, which inherently increases computational time.
Finally, it is not implemented in any readily available software.

Despite these advancements, these approaches were developed outside the traditional SEM framework.
Consequently, developments in the traditional SEM framework cannot be directly applied, such as conducting comprehensive model assessments, handling missing data, and model comparisons.

\subsection{Model Specifications to Accommodate Common Factors and Composites in the Traditional SEM Framework}

Next to approaches that emerged outside the traditional SEM framework, approaches have been developed that allow for integrating composites into the traditional SEM framework.
Consequently, these approaches can  deal with models that include both composites and common factors.

In the classical one-step approach \parencite{Grace2008}, a composite is modeled as a formatively measured latent variable whose error term variance is fixed to zero.
Consequently, the latent variable is entirely composed of its indicators, i.e., it is a composite.
This specification restricts composites to independent positions in the structural model and prohibits the estimation of covariances between the composite and other variables \parencite{Schuberth2023e}. 
This limitation can be partly addressed by modeling the effects of other variables on the composite via its indicators \parencite{Schamberger2025c}.
Yet, this significantly increases the number of model parameters and therefore reduces the degrees of freedom.

Another approach to study composites in SEM is the pseudo-indicator approach \parencite{Rose2019}.
This approach exploits the idea that instead of modeling a composite as a weighted linear combination of its indicators, one indicator can be expressed as the difference between the composite and other indicators.
However, this method currently lacks a mechanism for freely estimating composite weights.

The Henseler--Ogasawara (H--O) specification \parencite{Henseler2021, Schuberth2023e, Yu2023} overcomes the limitations of the pseudo-indicator and the one-step approach..
This specification extracts as many composites as there are indicators and expresses the relationships between composites and indicators using composite loadings.
Although promising, the H–O specification has several limitations. 
First, it increases specification complexity by introducing additional variables.
Second, it specifies composites in terms of composite loadings, which contradicts the definition of these constructs.
Consequently, the actual model parameters of interest, i.e., the composite weights, variances, and covariances between the corresponding indicators, must be derived from the estimated model parameters after estimation.
As a result, restricting composite weights is not straightforward \parencite{Henseler2025a} and even more challenging for composite indicators' variances and covariances.
Third, applying this specification in traditional SEM software requires the use of several tricks, such as fixing measurement error variances to zero and finding appropriate starting values.
Together, these points often lead to convergence issues, and increased computational time.

In conclusion, current approaches for handling composites and common factors in structural equation models face limitations regarding model specification, identification, and assessment.
To fully leverage the capabilities of traditional SEM, we need to address these limitations.

\section{Structural Equation Modeling with Common Factors and Composites}
\label{sec:SEM}

This section introduces FC-SEM, a new approach to incorporate common factors and composites in the traditional SEM framework.
Specifically, we present how to specify the model and we derive the model-implied variance-covariance matrix.
We further demonstrate how to ensure model identification.
Finally, we demonstrate how to estimate and assess the models with approaches developed in the traditional SEM framework.

\subsection{Model Specification}

SEM centers on the structural equation model, a statistical model that represents relationships among variables using linear structural equations.
In empirical studies, it is usually a formalized representation of researchers' conceptual understanding of the real world.
We will demonstrate how to specify structural equation models that incorporate both composites and common factors.
For simplicity, we assume that each variable has a mean of zero.
Any structural equation model can be divided into two components: (i) the models that contain relations between constructs and their indicators, and (ii) the model that relates the constructs, i.e., the structural model.
We will introduce both models in the following.

In the case of common factors and composites, we have two models that relate the constructs with their indicators as shown in the following equations:
\begin{align}
    \bm y^f &= \bm \Lambda^f \bm \eta^f + \bm \varepsilon^f \label{eq:LVmodel}\\
    \bm \eta^c &= \bm W' \bm y^c \label{eq:compositemodel}
\end{align}
where $\bm y^f$ is a ($P^f \times 1$) random vector of the indicators related to the common factors, which are comprised in the ($M^f \times 1$) random vector $\bm \eta^f$.
Consequently, $\bm y^f$ is referred to as \textit{common factor indicators}. 
Further, the ($P^f \times 1$) random vector $\bm \varepsilon^f$ comprises the random measurement errors of the common factor indicators, which are assumed to be uncorrelated with $\bm \eta^f$ and $\bm \eta^c$.
Similarly, $\bm y^c$ is a ($P^c \times 1$) random vector of the indicators related to the composites, which are comprised in the ($M^c \times 1$) random vector $\bm \eta^c$.
Thus, this type of indicators is referred to as \textit{composite indicators}.
The ($P^f \times M^f$) matrix $\bm \Lambda^f$ contains the effects of the common factors on their indicators, commonly referred to as factor loadings and the ($P^c \times M^c$) matrix $\bm W$ contains the contributions of the composite indicators to the composites, i.e., the composite weights.

While the variances and covariances among common factor indicators are explained by the common factor and the random measurement errors, the variances of the composite indicators are unrestricted, as are the covariances between indicators that form a specific composite.
These unrestricted variances and covariances are captured in the ($P^c \times P^c$) matrix $\bm T$ which is defined as follows:
\begin{align}
    T_{ij} = \begin{cases}
        \text{cov}(y^c_i, y_j^c)\quad & \text{if the two indicators }y^c_i\text{ and } y_j^c\text{ form the same composite}\\
        0 \quad & \text{otherwise.}
    \end{cases}
    \label{eq:T}
\end{align}

Next to the relations between the constructs and their indicators, constructs are allowed to be related with each other.
These relations are expressed in the structural model:
\begin{align}
     \bm \eta = \bm B \bm \eta + \bm \zeta \qquad \Leftrightarrow \qquad \bm \eta = (\bm I - \bm B)^{-1} \bm \zeta,
     \label{eq:structural}
\end{align}
where the $(M \times 1)$ vector $\bm \eta$ captures the common factors and composites, the ($M \times M$) matrix $\bm B$ contains the relations between the constructs, and the ($M \times 1$) random vector $\bm \zeta$ contains the structural error terms\footnote{If the construct $\eta_i$ is not explained by other variables in the structural model, the corresponding structural error term $\zeta_i$ is equal to the construct $\eta_i$. We further assume that $\bm I - \bm B$ has full rank.} which are assumed to be uncorrelated with $\bm \varepsilon^f$.
Moreover, relations between constructs and observed variables which do not serve as indicators for a common factor or a composite, i.e. covariates, can be modeled.
To do so, these observed variables, which are comprised in the random vector $\bm x$, can be modeled as single-indicator common factors where the indicators have no measurement error and the factor loadings are set to 1.
Consequently, they are just special cases of the common factor model and can also be subsumed in $\bm y^f$.

\subsection{Model Implications - The Model-Implied Variance-Covariance Matrix}

A core principle of traditional SEM is that indicator variances and covariances are modeled in terms of the model parameters.
This can be done by using implications of the model specification on the variables' variances and covariances.
Consequently, we derive the model-implied variance-covariance matrix of the indicators.

For common factor indicators as depicted in Equation (\ref{eq:LVmodel}), the model-implied variance-covariance matrix is well established \parencite[e.g.,][]{Bollen1989}:
\begin{align}
    \bm V(\bm y^f) = 
    \bm \Sigma(\bm \theta)^f = \bm \Lambda^f \text{V} (\bm \eta^f) \bm \Lambda^{l'} + \bm \Theta^f,
\end{align}
where $\bm \Theta^f$ represents the variance-covariance matrix of the measurement errors $\bm \varepsilon^f$ and $\text{V} (\bm \eta^f)$ represents the variance-covariance matrix of the common factors.

In the composite model, the variances and covariances of composite indicators forming a specific composite are free model parameters and are comprised in the matrix $\bm T$.
In contrast, the composites' variances are determined within the model and given by the diagonal elements of $\bm W' \bm T \bm W$.
To obtain the covariances between common factor indicators and composite indicators and the covariances between composite indicators which form different composites, we derive so called \textit{composite loadings}, i.e., coefficients from an ordinary least squares regression of the composites on their directly related indicators:
\begin{align}
    \bm y^c &= \bm \Lambda^c \bm \eta^c\\
\label{eq:compositeloading}
 \bm \Lambda^c &= \bm T \bm W (\bm W' \bm T \bm W)^{-1},
\end{align}
where $\bm T \bm W$ captures the covariances between the composites and their related indicators and the matrix $\bm W' \bm T \bm W$ is a diagonal matrix containing the composite variances.
We show the derivation of Equation (\ref{eq:compositeloading}) in the Appendix.
Consequently, the model-implied variance-covariance matrix for a model solely containing composites can be derived as:
\begin{align}
    \bm V(\bm y^c) = \bm \Sigma(\bm \theta)^c &= \bm \Lambda^c \text{V}(\bm \eta^c) \bm \Lambda^{c'} - \bm \Lambda^c \bm W'\bm T \bm W \bm \Lambda^{c'} + \bm T\\
    &=  \bm \Lambda^c \text{V}(\bm \eta^c) \bm \Lambda^{c'} + \bm \Theta^c.
\end{align}
Where the matrix $\bm \Theta^c$ is included to take into account that the variances of the composite indicators and covariances between composite indicators forming one specific composite are unrestricted and given by $\bm T$.
Note that in contrast to the common factor model, the matrix $\bm \Theta^c$ is not a variance-covariance matrix.

Further, using the relations of the structural model, see Equation (\ref{eq:structural}), the variance-covariance matrix of $\bm \eta$ is derived in the following equation:
\begin{align}
    \text{V}(\bm \eta) = (\bm I - \bm B)^{-1} \text{V}(\bm \zeta) (\bm I - \bm B)'^{-1}.
    \label{eq:constructVCV}
\end{align}
As the composites are dependent variables in the composite model, it needs to be taken into account that the variances of the composites are also determined via their relations with the indicators.
To combine the restrictions of the composite model and the structural model, the variances of the composites given by the last $M^c$ diagonal elements in Equation (\ref{eq:constructVCV}) must be set equal to the diagonal elements of $\bm W' \bm T \bm W$.
Consequently, the variances of the structural error terms related to composites are not free parameters, but functions of the remaining model parameters. 

If we combine the constraints of all parts of the model, we can derive the model-implied variance-covariance matrix of the indicators.
To do so, we make use of the following equation:
\begin{align}
    \begin{pmatrix}
        \bm y^f\\
        \bm y^c
    \end{pmatrix} = \begin{pmatrix}
        \bm \Lambda^f & \bm 0 \\
        \bm 0 & \bm \Lambda^c
    \end{pmatrix} \begin{pmatrix}
        \bm \eta^f\\
        \bm \eta^c 
    \end{pmatrix} + \begin{pmatrix}
        \bm \varepsilon^f \\
        \bm 0
    \end{pmatrix} \label{eq:compositelvmodel}
    \end{align}
By combining $\bm y^f$ and $\bm y^c$ into a single vector, i.e. $\bm y = (\bm y^f, \bm y^c)'$, we can write this more compactly as:    
\begin{align}
    \bm y = \bm \Lambda \bm \eta + \bm \varepsilon
\end{align}
Subsequently,  the model-implied variance-covariance matrix of the indicators is given by:
\begin{align}
    \bm \Sigma(\bm \theta) = \bm \Lambda \text{V} (\bm \eta) \bm \Lambda ' + \bm \Theta,
    \label{eq:Sigma}
\end{align}
where we define $\bm \Theta$ as:\footnote{Note that covariances between composite indicators and measurement errors could in principle be defined via the off-diagonal blocks.}
$\bm \Theta = \begin{pmatrix}
    \bm \Theta^f & \bm 0\\
    \bm 0 & \bm \Theta^c
\end{pmatrix}$ with $\bm \Theta^c = \bm T - \bm \Lambda^c \bm W'\bm T \bm W \bm \Lambda^{c'}$.
Note that the model-implied variance-covariance matrix strongly resembles the one known from the traditional SEM framework.

\subsection{Model Identification}
In this section, we present the necessary conditions for identifying a structural equation model with both composites and common factors.
Ensuring that the model parameters are identified is crucial because otherwise the results cannot be interpreted reliably, as re-estimating the same model with identical data can produce entirely different results \parencite{Bollen1989}.

The first necessary condition for identification is adhering to the $t$ rule, which requires that the degrees of freedom are greater than or equal to zero \parencite{Bollen2009}.
The degrees of freedom are defined as the difference between the number of the non-redundant variances and covariances of the indicators and the number of the model parameters $K$:
\begin{align}
    \text{df} = (P \cdot (P + 1))/2 - K
    \label{eq:DF}
\end{align}
For a model with both composites and common factors, the model parameters are given by:
\begin{itemize}
    \item The non-redundant elements in $\bm \Theta^f$, i.e., the free (co)variances of the measurement errors
    \item The non-redundant elements in $\bm T$, i.e., the free (co)variances of the composite indicators
    \item The free factor loadings in $\bm \Lambda^f$
    \item The free composite weights in $\bm W$
    \item The free directed relations between constructs in $\bm B$
    \item The free variances of the structural error terms, i.e., the respective $M^f$ main diagonal elements in $V(\bm \zeta)$
    \item The free covariances of the structural error terms, i.e., the off-diagonal elements in $V(\bm \zeta)$
\end{itemize}

Ensuring model identification further requires that all sub-models -- common factor model, composite model, and structural model -- are identified.
For the structural model, traditional SEM rules apply.
We refer to \textcite{Bollen2009} and \textcite{Obrien1994} for further discussion.

Similarly, we can rely on the rules known from the traditional SEM framework for identification of the common factor model.
Scaling each common factor is crucial.
For this purpose, several scaling methods have been proposed \parencite[e.g.,][]{Klopp2021}, e.g., fixing one loading per common factor, or fixing the variances of the common factors.
Additional identification rules are discussed in \textcite{Bollen1989}.

For the composite model, scaling of the composites is necessary. 
Our model permits alternative scaling methods.
Specifically, composites can be scaled by fixing one weight, fixing the variances of composites, or imposing other restrictions on the weights, such as constraining their sum. 
Additionally, a composite needs to be related to at least one other variable not forming the composite \parencite{Dijkstra2017}.
This is because the variances and covariances of observed variables making up a composite are free parameters, and thus, an isolated composite would have a negative number of degrees of freedom unless all weights are pre-specified.

To conclude, identifying a structural equation model that includes both composites and common factors involves applying traditional SEM rules for common factor and structural models and ensuring that the composite model is identified through proper scaling and relating the composites to other variables (or fixing all composite weights).

\subsection{Model Estimation}
Once a structural equation model is specified and all parameters are identified, we can proceed to estimate the model parameters. 
Our proposed model specification, along with its model-implied variance-covariance matrix, enables us to directly apply established methods from the traditional SEM framework for this purpose.
Consequently, estimation of structural equation models involving composites and common factors is conducted in the same manner as the estimation of structural equation models with common factors only.

In the traditional SEM framework, the model parameters are estimated by minimizing the discrepancy between the model-implied (see Equation \ref{eq:Sigma}) and the sample variance-covariance matrix of the indicators.
For this purpose, various optimization criteria, such as weighted least squares \parencite[WLS,][]{Browne1984}, generalized least squares \parencite[GLS,][]{Joereskog1972}, two-stage least squares \parencite[2SLS,][]{Bollen1996}, and ML \parencite{Joereskog1969, Joereskog1970} can be used. 
ML is the most widely used approach, likely due to its capabilities, such as handling missing values, and the fact that it is the default method in most SEM software packages.
Therefore, we show in the following how to obtain ML parameter estimates for our presented model specification.
For this purpose, we additionally assume that the observed variables are normally distributed, i.e., $\bm y \sim N(\bm 0, \bm \Sigma (\bm \theta))$.
The ML estimates are obtained by maximizing the log likelihood function \parencite{Lawley1963}:
\begin{align}
\text{log L} = -\frac{1}{2}n \left[ \text{log}| \bm \Sigma(\bm \theta)| + \text{tr}(\bm S \bm \Sigma(\bm \theta)^{-1}) \right] - \frac{N \cdot P}{2} \cdot \text{log}(2 \cdot \pi)
\end{align}
Maximizing the log likelihood function is equivalent to minimizing the following discrepancy function:
\begin{align}
\label{eq:F}
\text{F} = \text{log}|\bm \Sigma(\bm \theta)| + \text{tr}(\bm S \bm \Sigma(\bm \theta)^{-1}) - \text{log}|\bm S| - P,
\end{align}
where the constants log$|\bm S|$ and $P$ are included to obtain a value of $0$ for F if $\bm \Sigma(\hat{\bm \theta})$ equals $\bm S$.
The results for the remaining estimators can be obtained by optimizing their respective criterion.

\subsection{Model Assessment}

Employing traditional SEM estimators, including ML, WLS, and GLS, not only provides consistent, asymptotically unbiased and efficient estimates \parencite{Davidson1993}, but also provides various model assessment options.
Specifically, all analytic robust and non-robust standard errors and test statistics developed in the traditional SEM framework \parencite{Savalei2021} can be used for assessing individual parameter estimates.

In addition to assessing individual parameter estimates, our approach allows us to use traditional SEM methods to assess the overall model fit and determine whether the model is consistent with the collected data.
For example, the $\chi^2$ test \parencite{Wilks1938, Joereskog1969} can be used to test the following null hypothesis:
\begin{align}
H_0: \bm \Sigma(\bm \theta) = \bm \Sigma
\end{align}
where  $\bm \Sigma$ is the population variance-covariance matrix of the observed variables $\bm y$. 
The test statistic is obtained as follows \parencite{Wilks1938,Joereskog1969}: 
\begin{align}
\chi^2 = n \cdot F(\hat{\bm \theta}), 
\end{align}
where  $F$ is the fit function and the sample size is denoted by $n$.
This test statistic is under the null hypothesis of perfect fit asymptotically $\chi^2$ distributed with $df$ degrees of freedom as given in Equation (\ref{eq:DF}). 

The $\chi^2$ test can be used to test exact fit.
To evaluate close fit instead, researchers can use fit indices such as the standardized root mean squared residual \parencite[SRMR,][]{Joereskog1988} or the root mean square error of approximation \parencite[RMSEA,][]{Brown1993}.
For a detailed overview of these fit indices and their interpretation, we refer to \textcite{Hu1999} and \textcite{Schermelleh-Engel2003}.

\section{Implementation in lavaan}

In this section, we explain the implementation of FC-SEM in the open-source software R \parencite{RCT2020} package lavaan \parencite[version 0.6-20][]{Rosseel2012, Rosseel2025}.
This package is very popular in the field of SEM and is frequently used in practice \parencite{Jorgensen2017}.
Generally, the analysis of structural equation models involving composites and common factors follows a procedure similar to the analysis of models involving only common factors.
Specifically, the same functions are applicable, and there is no need to modify the model syntax for the structural and common factor models.
In the following, we briefly outline the necessary adaptations for integrating composites.

To incorporate composites in the lavaan syntax, the operator `$<\sim$' is used, with the composite specified on the left side and the composite indicators connected via `+' operators.\footnote{Note that in earlier package versions, the $<\sim$ operator was used to specify a formatively measured latent variable whose error term variance is fixed to zero.}
By default, composites are scaled, setting the weight of the first indicator per composite to one.
This is similar to the default scaling method for common factors.

The \texttt{sem()} function can be used to obtain the parameter estimates. 
To obtain the standardized parameter estimates, the \texttt{standardized} argument can be set to `TRUE' in the \texttt{summary()} function. 
Standardized parameter estimates are particularly useful for comparisons with PLSc results, as PLSc estimates are usually standardized. 
By default, the composite indicator (co)variances are fixed to their sample statistics. 

The current implementation (lavaan 0.6-20) has some limitations.
Firstly, while standardized parameter estimates can be obtained, scaling constructs to have unit variance during the estimation procedure is not yet supported when composites are involved.
Secondly, due to the absence of an analytical gradient, the option \texttt{optim.gradient} must be set to \texttt{"numerical"} during optimization.
Lastly, the current implementation does not yet support mean structures.
Nevertheless, these limitations are expected to be addressed in future updates.

\section{Scenario Analysis And Finite Sample Performance}
\label{sec:scenario}

In this section, we demonstrate our approach to SEM with common factors and composites through a scenario analysis.
Specifically, we illustrate how to specify the model, how to derive the model-implied variance-covariance matrix, how to identify the model parameters, and how to calculate the degrees of freedom.
Moreover, we demonstrate that our approach can retrieve the population parameters when the data perfectly represents the population.
In this case, evaluating the overall model fit is not informative, as it would inevitably indicate a perfect fit.
Finally, we evaluate the finite sample behavior of our approach.\footnote{The R code is available via the following link: \url{https://osf.io/9tm2y/?view_only=59d9792bab994892a735e4efc763b511}}

\begin{figure}[H]
    \centering
        \caption{Population Model}
    \label{fig:popmodel}
    \includegraphics[width=1\linewidth]{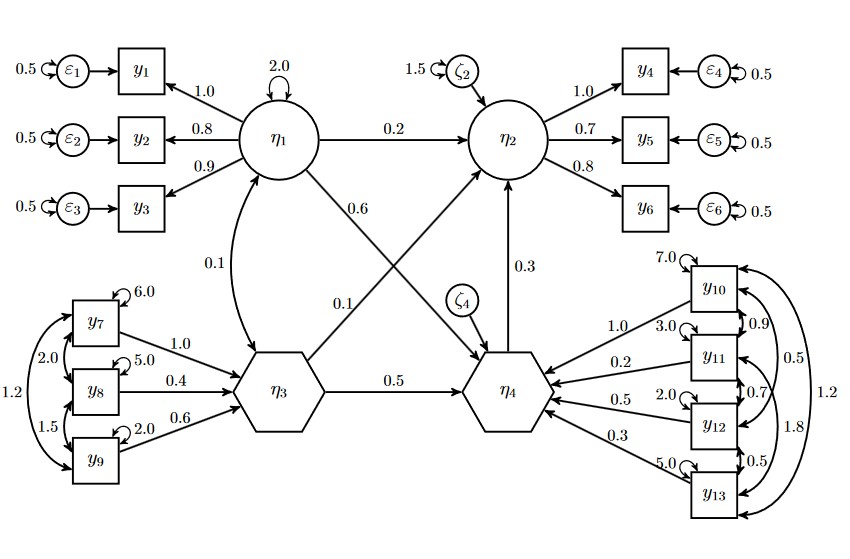}
\end{figure}

We consider a population model with four constructs $\eta_1$, $\eta_2$, $\eta_3$, and $\eta_4$.
The constructs $\eta_1$ and $\eta_2$ are modeled as common factors, while the constructs $\eta_3$ and $\eta_4$ are modeled as composites.
With this population model, our approach can be illustrated in the most general, but also most simple case.
The population model is illustrated in Figure \ref{fig:popmodel}.

The two common factors $\eta_1$ and $\eta_2$ are measured by three observed variables:
\begin{align}
    y_{1} &= 1 \cdot \eta_1 + \varepsilon_1; \qquad y_{2} = 0.8 \cdot \eta_1 +\varepsilon_2; \qquad y_3 = 0.9 \cdot \eta_1 + \varepsilon_3 \\
    y_4 &= 1 \cdot \eta_2 + \varepsilon_4; \qquad y_5 = 0.7 \cdot \eta_2 + \varepsilon_5; \qquad y_6 = 0.8 \cdot \eta_2 + \varepsilon_6.
\end{align}
The measurement errors are assumed to be uncorrelated with a variance of 0.5 each.
Additionally, the random measurement errors are assumed to be uncorrelated with the common factors and the composites.
The composites $\eta_3$ and $\eta_4$ are related to their indicators as follows:
\begin{align}
    \eta_3 &= 1 \cdot y_7 + 0.4\cdot  y_8 + 0.6 \cdot  y_9 \\
    \eta_4 &= 1 \cdot y_{10} + 0.2 \cdot y_{11} + 0.5 \cdot y_{12} + 0.3 \cdot y_{13}.
\end{align}
The variance-covariance matrices of the indicators forming a composite are given by:
\begin{align}
    \bm V(y_7, y_8, y_9) &= \begin{pmatrix}
        6.0&  2.0& 1.2 \\
        2.0& 5.0& 1.5 \\
        1.2& 1.5& 2.0
    \end{pmatrix}; \qquad \bm V(y_{10}, y_{11}, y_{12}, y_{13}) = \begin{pmatrix}
        7.0& 0.9& 0.5& 1.2\\ 
        0.9& 3.0& 0.7& 1.8\\
        0.5& 0.7& 2.0& 0.5\\
        1.2& 1.8& 0.5& 5.0\\
    \end{pmatrix}.
\end{align}
The structural model is specified as: 
\begin{align}
    \eta_4 &= 0.6\cdot \eta_1 + 0.5\cdot\eta_3 + \zeta_4 \\
    \eta_2 &= 0.2 \cdot\eta_1 + 0.1\cdot\eta_3 + 0.3 \cdot\eta_4 + \zeta_2.
\end{align}
Here, $\zeta_2$ and $\zeta_4$ are uncorrelated, uncorrelated with $\eta_1$, and $\eta_3$, and $\zeta_2$ has a variance of 1.5.
The variance of $\eta_1$ is set to 2.0, and the covariance between $\eta_1$ and $\eta_3$ is 0.1.
The population variance-covariance matrix and its derivation is provided in the Appendix.  

To evaluate the performance of our approach in the case where the sample perfectly represents the population, we draw a sample of 200 observations that exactly represents this population, i.e., the sample variance-covariance matrix of the observed variables equals the population counterpart, from the multivariate normal distribution using the R package lavaan.
Subsequently, we specify the common factor model as:
\begin{align*}
    \begin{pmatrix}
        y_1 \\
        y_2 \\
        y_3 \\
        y_4 \\
        y_5 \\
        y_6
    \end{pmatrix}&= \begin{pmatrix}
        \lambda_1 & 0\\
        \lambda_2 & 0 \\
        \lambda_3 & 0 \\
        0 & \lambda_4 \\
        0 & \lambda_5 \\
        0 & \lambda_6
    \end{pmatrix} \begin{pmatrix}
        \eta_1 \\
        \eta_2
    \end{pmatrix}  + \begin{pmatrix}
        \varepsilon_1 \\
        \varepsilon_2 \\
        \varepsilon_3 \\
        \varepsilon_4 \\
        \varepsilon_5 \\
        \varepsilon_6 
    \end{pmatrix}.
\end{align*}
The composite model is specified as:
\begin{align*}
\begin{pmatrix}
    \eta_3 \\
    \eta_4
\end{pmatrix} &= \begin{pmatrix}
    w_7 & w_8 & w_9 & 0 & 0 & 0 & 0\\
    0 & 0 & 0 & w_{10} & w_{11} & w_{12} & w_{13}
\end{pmatrix} \begin{pmatrix}
        y_7 \\
        \vdots \\
        y_{13}
    \end{pmatrix}.
\end{align*}
The matrix $\bm T$ is given by: 
\begin{align}
    \bm T &= \begin{pmatrix}
        \text{var}(y_7)  \\
        \text{cov}(y_7, y_8) & \text{var}(y_8)\\
        \text{cov}(y_7, y_9) & \text{cov}(y_8, y_9) & \text{var}(y_9)\\
        0 & 0 & 0 & \text{var}(y_{10})\\
        0 & 0 & 0 & \text{cov}(y_{10}, y_{11}) & \text{var}(y_{11}) \\
        0 & 0 & 0 & \text{cov}(y_{10}, y_{12}) & \text{cov}(y_{11}, y_{12}) & \text{var}(y_{12})\\
        0 & 0 & 0 & \text{cov}(y_{10}, y_{13}) & \text{cov}(y_{11}, y_{13}) & \text{cov}(y_{12}, y_{13}) & \text{var}(y_{13})\end{pmatrix}.
\end{align}
Moreover, the matrix $\bm \Theta^f$ contains the variances $\theta_{ii}$ of measurement errors $\varepsilon_i$ on the diagonal.
The variances of the composites are calculated as:
\begin{align}
    \text{var}(\eta_3) &= \begin{pmatrix}
        w_7 & w_8 & w_9
    \end{pmatrix} \begin{pmatrix}
        \text{var}(y_7)  \\
        \text{cov}(y_7, y_8) & \text{var}(y_8)\\
        \text{cov}(y_7, y_9) & \text{cov}(y_8, y_9) & \text{var}(y_9)
    \end{pmatrix} \begin{pmatrix}
        w_7 \\ w_8 \\ w_9
    \end{pmatrix} \label{eq:vareta3} \\ 
    \text{var}(\eta_4) &= \begin{pmatrix}
        w_{10} & w_{11} & w_{12} & w_{13}
    \end{pmatrix}\begin{pmatrix}
      \text{var}(y_{10})\\
        \text{cov}(y_{10}, y_{11}) & \text{var}(y_{11}) \\
         \text{cov}(y_{10}, y_{12}) & \text{cov}(y_{11}, y_{12}) & \text{var}(y_{12})\\
        \text{cov}(y_{10}, y_{13}) & \text{cov}(y_{11}, y_{13}) & \text{cov}(y_{12}, y_{13}) & \text{var}(y_{13})  
    \end{pmatrix}\begin{pmatrix}
        w_{10} \\ w_{11} \\ w_{12} \\ w_{13}
    \end{pmatrix} .
    \label{eq:varcomp}
\end{align}
The structural model is expressed as:
\begin{align}
    \begin{pmatrix}
        \eta_1 \\
        \eta_2 \\
        \eta_3 \\
        \eta_4
    \end{pmatrix} &= \begin{pmatrix}
        0 & 0 & 0 & 0 \\
        b_3 & 0 & b_4 & b_5 \\
        0 & 0 & 0 & 0\\
        b_1 & 0 & b_2 & 0
    \end{pmatrix} \begin{pmatrix}
        \eta_1 \\
        \eta_2 \\
        \eta_3 \\
        \eta_4
    \end{pmatrix} + \begin{pmatrix}
        \eta_1 \\
        \zeta_2 \\
        \eta_3 \\
        \zeta_4
    \end{pmatrix}.
\end{align}
The variance of $\eta_1$ is denoted as $\psi_{11}$, the variance of $\zeta_2$ is $\psi_{22}$, the variance of $\eta_3$ is given by Equation (\ref{eq:vareta3}), the covariance between $\eta_1$ and $\eta_3$ is denoted as $\psi_{13}$, and the variance of $\zeta_4$ is calculated as:
\begin{align}
    \text{var}(\zeta_4) = \psi_{44} = \text{var}(\eta_4) - b_1^2 \text{var}(\eta_1) - b_2^2 \text{var}(\eta_3) - 2b_1 b_2 \text{cov}(\eta_1, \eta_3).
    \label{eq:varzeta4}
\end{align}
The model-implied variance-covariance matrix of the indicators can be derived using the Equations (\ref{eq:constructVCV}) and (\ref{eq:Sigma}).

To ensure model identification, we scale the common factors and composites by setting $\lambda_1$, $\lambda_4$, $w_7$, and $w_{10}$ to 1. 
The common factor model is identified as each common factor is scaled and fulfills the two-indicator rule, i.e., each common factor is related to at least two indicators and the measurement errors are uncorrelated.
The composite model is identified since each composite is scaled and related to at least one other variable next to its indicators. 
The structural model is identified as it meets the recursive rule, i.e., the structural model contains no feedback loops and the structural error terms are uncorrelated and uncorrelated with $\eta_1$ and $\eta_3$. 
Furthermore, no indicator is related to more than one construct.
Following this, we calculate the degrees of freedom.
It is important to note that, unlike common factors, the variances of composites are inherently determined within the model.
This applies to composites in both dependent and independent positions in the structural model.
Consequently, as shown in Equations (\ref{eq:vareta3}) and (\ref{eq:varcomp}), the variances of the composites depend on the other model parameters. 
As a result, the variances of the structural error terms related to composites, such as $\zeta_4$ in this example, are fully determined by the remaining model parameters, as illustrated in Equation (\ref{eq:varzeta4}).
The degrees of freedom of our example model are calculated as follows:
\begin{align*}
    \text{df} =& \underbrace{91}_{\shortstack{\tiny \text{Number of non-redundant variances and}\\ \tiny\text{covariances of the indicators}}} - \underbrace{6}_{\shortstack{\tiny \text{Number of free variances}\\\tiny\text{of the measurement errors}}}- \underbrace{16}_{\shortstack{\tiny \text{Number of free variances and}\\ \tiny \text{covariances of the composite indicators}}} - \underbrace{4}_{\shortstack{\tiny \text{Number of free latent}\\ \tiny \text{variable loadings}}} -\\& \underbrace{5}_{\shortstack{\tiny \text{Number of free}\\ \tiny \text{composite weights}}} - \underbrace{5}_{\shortstack{\tiny \text{Number of free} \\ \tiny \text{elements in $\bm B$}}} - \underbrace{1}_{\shortstack{\tiny \text{Number of free variances}\\ \tiny \text{of structural error terms}}} -  \underbrace{1}_{\shortstack{\tiny \text{Number of free}\\ \tiny \text{construct covariances}}} - \underbrace{1}_{\shortstack{\tiny \text{Number of free variances} \\ \tiny \text{of common factors}}} = 52 > 0.
\end{align*}

Once the model is identified, we can estimate and asses it using the methods known from the traditional SEM framework.
To do so, we make use of the R package lavaan as explained in the previous section.
Specifically, we estimate the parameters using the ML estimator.
Table \ref{tab:scenario_estimates} shows the parameter estimates.
Given that the data perfectly represents the population, all population parameters were retrieved.

\begin{table}[h]
    \centering
      \caption{Parameter Estimates Obtained by FC-SEM for a Sample Perfectly Representing the Population}
    \begin{tabular}{ccc}
    \toprule
    Parameter & Population value & Estimate\\ \midrule
       $\lambda_2$  &  0.8 & 0.8\\
       $\lambda_3$  &  0.9 & 0.9\\
       $\lambda_5$ & 0.7 & 0.7\\
       $\lambda_6$ & 0.8 & 0.8\\
       $w_8$ & 0.4 & 0.4\\
       $w_9$ & 0.6 & 0.6\\
       $w_{11}$ & 0.2 & 0.2\\
       $w_{12}$ & 0.5 & 0.5\\
       $w_{13}$ & 0.3 & 0.3\\
       $b_1$ & 0.5 & 0.5\\
       $b_2$ & 0.6 & 0.6\\
       $b_3$ & 0.2 & 0.2\\
       $b_4$ & 0.3 & 0.3\\
       $b_5$ & 0.1 & 0.1\\
       $\psi_{11}$ & 2.0 & 2.0\\
       $\psi_{22}$ & 1.5 & 1.5\\
       $\psi_{13}$ & 0.1 & 0.1\\
       $\theta_i$ ($i = 1,\dots 6$) & 0.5 & 0.5 \\
       \bottomrule
    \end{tabular}
    \label{tab:scenario_estimates}
\end{table}

Next, we evaluate the finite sample behavior of our approach.
FC-SEM enables us to use estimators from the traditional SEM framework to estimate model parameters.
Consequently, we do not expect the finite sample behavior of the ML estimator, for example, to be different when applied to models involving composites and common factors than when applied to models involving only common factors.
To verify this further, we ran a small Monte Carlo simulation in which we drew 1,000 samples from a multivariate normal distribution with a mean vector of zero and the population variance-covariance matrix of the indicators, each with a sample size of 500.
Table (\ref{tab:simulation}) illustrates the average parameter estimates and the ratio between the average standard errors and the empirical standard deviations of the point estimates across all replications.
As expected, the parameter estimates are, on average, close to their population counterparts.
Moreover, most of the average standard errors are close to the empirical standard errors.
In summary, our approach yields consistent estimates that are, on average, close to their population counterparts. 

\begin{table}[h]
    \centering
      \caption{Average Parameter Estimates and Ratio Between The Average Standard Errors and the Empirical Standard Deviations Obtained by FC-SEM for the Simulation Study (N = 500)}
    \begin{tabular}{cccc}
    \toprule
    Parameter & Population Value & Estimate & $\frac{\text{Average Standard Error}}{\text{Empirical Standard Deviation}}$\\ \midrule
       $\lambda_2$  &  0.8 & 0.798 & 0.905\\
       $\lambda_3$  &  0.9 & 0.900 & 0.919\\
       $\lambda_5$ & 0.7 & 0.699 & 0.782\\
       $\lambda_6$ & 0.8 & 0.800 & 0.813\\
       $w_8$ & 0.4 & 0.403 & 0.933\\
       $w_9$ & 0.6 & 0.609 & 0.926\\
       $w_{11}$ & 0.2 & 0.198 & 0.985\\
       $w_{12}$ & 0.5 & 0.496 & 0.983\\
       $w_{13}$ & 0.3 & 0.306 & 0.987\\
       $b_1$ & 0.5 & 0.505& 0.984\\
       $b_2$ & 0.6 & 0.604 & 1.004\\
       $b_3$ & 0.2 & 0.200 & 1.118\\
       $b_4$ & 0.3 & 0.300 & 1.717\\
       $b_5$ & 0.1 & 0.100 & 1.443\\
       $\psi_{11}$ & 2.0 & 1.998 & 1.015\\
       $\psi_{22}$ & 1.5 & 1.490 & 1.149\\
       $\psi_{13}$ & 0.1 & 0.094& 1.004\\
       \bottomrule
    \end{tabular}
    \label{tab:simulation}
\end{table}

\section{Empirical Example}
\label{sec:empirical}
In this section, we demonstrate how FC-SEM can be employed in the R package \texttt{lavaan} by means of an empirical example.
Additionally, we compare the results of our approach to those of IGSCA, PLSc, and the H--O specification. 
For this purpose, we employ the American Customer Satisfaction Index dataset, which has recently been examined in the context of IGSCA \parencite{Hwang2023} and PLS-PM \parencite{Hair2022}.
The model comprises eight constructs in total, four of which are modeled as composites: quality, composed of eight indicators; performance, composed of five indicators; corporate social responsibility, composed of five indicators; and attractiveness, composed of three indicators. 
The remaining four constructs are modeled as common factors: competence, customer loyalty, and likeability, each with three measures, and customer satisfaction, which is measured by a single indicator.
For a detailed description of the data, we refer the reader to \textcite{Hair2022}.\footnote{The R Code used for the empirical example is available via the following link: \url{https://osf.io/9tm2y/?view_only=59d9792bab994892a735e4efc763b511}. The data is available via \textcite{Hwang2023}.}

The first step involves specifying the model in \texttt{lavaan} syntax as illustrated as follows:
\begin{lstlisting}[language=Octave]
model <- '
# Specification of the composites
Quality        <~ qual_1 + qual_2 + qual_3 + qual_4 +
                  qual_5 + qual_6 + qual_7 + qual_8
Performance    <~ perf_1 + perf_2 + perf_3 + perf_4 + perf_5
CorpSocResp    <~ csor_1 + csor_2 + csor_3 + csor_4 + csor_5
Attractiveness <~ attr_1 + attr_2 + attr_3

# Specification of the latent variables
Competence       =~ comp_1 + comp_2 + comp_3
CustomerLoyalty  =~ cusl_1 + cusl_2 + cusl_3
Likeability      =~ like_1 + like_2 + like_3

# Specification of the single indicator latent variable
CustomerSatisfaction =~ cusa

# Specification of the structural model
Competence           ~ Quality + Performance + CorpSocResp +
                       Attractiveness
Likeability          ~ Quality + Performance + CorpSocResp + 
                       Attractiveness
CustomerSatisfaction ~ Competence + Likeability
CustomerLoyalty      ~ CustomerSatisfaction + Competence + Likeability
'
\end{lstlisting}
Similar to scaling common factors, the default behavior in \texttt{lavaan} is to scale composites by setting the weight of their first indicator to one.

To conduct our approach to SEM with composites and common factors and estimate the model parameters, we utilize the \texttt{sem()} function of the \texttt{lavaan} package:
\begin{lstlisting}[language=Octave]
fit <- sem(model, data = data, optim.gradient = "numerical")
summary(fit, standardized = TRUE, fit.measures = TRUE)
\end{lstlisting}
Per default, the ML estimator is used to obtain parameter estimates.
Table \ref{tab:empirical_results} shows the standardized parameter estimates produced by our approach.
For simplicity, we only report the parameter estimates of the loadings of the indicators related to likeability, the weights to form attractiveness, and the structural model for customer loyalty.
The remaining model parameters along with the summary output can be found in the supplementary material.
Additionally, we report the results of IGSCA \parencite[using GSCAPro,][]{Hwang2021GSCAPRO}, PLSc \parencite[using the R package cSEM,][]{Rademaker2020} and the H--O specification (using the ML estimator as implemented in the package lavaan).
\begin{table}[h]
    \centering
    \caption{Standardized Parameter Estimates Obtained by FC-SEM, IGSCA, PLSc, and the H--O Specification for the Customer Satisfaction Example}
    \label{tab:empirical_results}
    \resizebox{1\textwidth}{!}{%
   \begin{tabular}{lccccccccccccc}
    \toprule
    \multirow{3}{87pt}{Parameter}  & \multirow{3}{87pt}{\centering FC-SEM (ML)} &  \multirow{3}{87pt}{\centering IGSCA} & \multirow{3}{87pt}{\centering PLSc}  & \multirow{3}{87pt}{\centering H--O Specification (ML)}\\   
     &&&&&\\
     &&&&&\\
    \midrule          
Likeability $\rightarrow$ like$_1$ & \phantom{-}0.833 & \phantom{-}0.801 & \phantom{-}0.857 & \phantom{-}0.833\\
       Likeability $\rightarrow$ like$_2$ & \phantom{-}0.774 & \phantom{-}0.811 & \phantom{-}0.763 & \phantom{-}0.774 \\
       Likeability $\rightarrow$ like$_3$ & \phantom{-}0.746 & \phantom{-}0.760 & \phantom{-}0.741 & \phantom{-}0.746\\
       attr$_1$ $\rightarrow$ Attractiveness & \phantom{-}0.430 & \phantom{-}0.512 & \phantom{-}0.418 & \phantom{-}0.430\\ 
       attr$_2$ $\rightarrow$ Attractiveness & \phantom{-}0.174 & \phantom{-}0.264 & \phantom{-}0.204 & \phantom{-}0.174 \\
       attr$_3$ $\rightarrow$ Attractiveness & \phantom{-}0.663 & \phantom{-}0.533 & \phantom{-}0.656 & \phantom{-}0.662\\
       CustomerSatisfaction $\rightarrow$ CustomerLoyalty & \phantom{-}0.488 & \phantom{-}0.535 & \phantom{-}0.501 & \phantom{-}0.488\\
       Competence $\rightarrow$ CustomerLoyalty & -0.090 & -0.102 & -0.115 & -0.090\\
       Likeability $\rightarrow$ CustomerLoyalty  & \phantom{-}0.492 & \phantom{-}0.464 & \phantom{-}0.530 & \phantom{-}0.491\\
     \bottomrule    
    \end{tabular}%
    }
\end{table}

Table \ref{tab:empirical_results} shows that the standardized parameter estimates are similar among the different approaches.
While the results of PLSc and IGSCA differ slightly from those of our approach due to the different estimation algorithms, the results of the H--O specification are almost identical to the results of our approach.
To conclude, the results of our proposed approach to SEM with common factors and composites are similar to those of other approaches to study such models, enabling researchers to draw comparable conclusions.

\section{Discussion}
\label{sec:discussion}
In this study, we introduce an approach to study structural equation models involving composites and common factors which allows us to derive a closed form expression for the variance-covariance matrix implied by such models.
Additionally, we present the necessary conditions for identifying the model parameters.
This specification along with its model-implied variance-covariance matrix allow us to apply approaches to model estimation and assessment developed in the traditional SEM framework.
We refer to our approach as factor- and composite-based SEM, i.e., FC-SEM.

We demonstrate the necessity of our approach by addressing the limitations of existing methods used to analyze structural equation models incorporating both common factors and composites.
Specifically, our approach overcomes the limitations of approaches developed outside the traditional SEM framework, such as PLSc, IGSCA, SVD-SEM, and RML-SEM.
First, unlike PLSc, our approach allows us to restrict model parameters, enabling us to incorporate prior knowledge.
Second, unlike PLSc, IGSCA, and SVD-SEM, our approach allows for the analytic derivation of standard errors, so inference on the model parameter estimates does not have to rely on non-parametric methods such as bootstrap.
Therefore, our approach can decrease computational time and enables the analytic derivation of the corresponding statistical properties.
Third, our approach enables the estimation of both standardized and unstandardized parameter estimates, as composites are not restricted to having a unit variance, overcoming limitations in PLSc, SVD-SEM, and RML-SEM.
Fourth, our approach allows for different scaling methods, so it does not rely on constrained optimization, as is done in RML-SEM.
Finally, our approach enables researchers to apply developments of the traditional SEM framework, such as model fit assessment and handling missing values, directly to models involving common factors and composites.
In contrast, such enhancements require additional effort for PLSc, IGSCA, SVD-SEM, and RML-SEM as researchers must develop new methods for these estimators instead of relying on existing methods.

Moreover, our approach overcomes the limitations of model specifications for structural equation models involving common factors and composites designed so that they can be incorporated in the traditional SEM framework.
Specifically, it overcomes the limitations of the one-step approach, the pseudo-indicator approach, and the H-O specification.
First, our approach addresses the shortcomings of the one-step approach by allowing composites to be included in dependent and independent positions of the structural model.
Consequently, composites can also be explained by other constructs, and particularly mediation analysis with composites as mediators or outcome variables is feasible.
Second, we overcome the limitations of the pseudo-indicator approach by allowing composite weights to be freely estimated.
Thus, weights can be obtained optimally within the model instead of outside the model framework.
Finally, FC-SEM overcomes the H-O specification's limitations in terms of model specification.
Specifically, with the H-O specification, researchers must specify the composites in terms of composite loadings by introducing additional variables.
Furthermore, to apply the H-O specification using traditional SEM software, researchers must restrict parameters, such as measurement errors related to composite indicators, to zero and adjust the starting values for model parameters related to the additional variables.
Additionally, due to the more complicated model specification, traditional SEM software tends to experience more convergence issues when the H-O specification is employed \parencite{Schamberger2025c}.
Finally, the actual parameters of a composite model -- i.e., the composite weights, indicator variances, and indicator covariances -- are no direct model parameters in the H-O specification. 
Rather, they must be derived from the model parameters obtained by this specification.
This complicates the inference and restriction of these parameters.
In summary, our approach surpasses existing methods for analyzing structural equation models involving composites and common factors in terms of model specification, identification, assessment, and application of traditional SEM approaches.
Further, our approach offers more flexibility in terms of model specification by, e.g., allowing for specifying cross-weights or covariances between composite indicators and measurement errors.
A detailed comparison of the capabilities of the different approaches is given in Table \ref{tab:Discussion}.

    \begin{adjustbox}{max width=\textwidth}
\begin{sidewaystable} 
\scriptsize
\begin{spacing}{1} 
    \centering
    \caption{Comparison of Various Approaches to Deal With Common Factors and Composites in SEM}
    \label{tab:Discussion}
    \begin{tabular}{p{7cm}p{1.5cm}p{1.5cm}p{1.5cm}p{1.5cm}p{1.5cm}p{1.5cm}p{1.5cm}p{1.5cm}}
        \toprule
     Capabilities of the different approaches &\multirow{3}{1.5cm}{\centering PLSc}& \multirow{3}{1.5cm}{\centering IGSCA} & \multirow{3}{1.5cm}{\centering SVD-SEM} & \multirow{3}{1.5cm}{\centering RML}  & \multirow{3}{1.5cm}{\centering One-Step Approach}& \multirow{3}{1.5cm}{\centering Pseudo-Indicator Approach}& \multirow{3}{1.5cm}{\centering H-O Specification}  & \multirow{3}{1.5cm}{\centering FC-SEM} \\ \\ \\
        \midrule
        \multirow{2}{8cm}{\textit{Model Specification}}\\ \\
        \multirow{2}{8cm}{\hphantom{000}Fix model parameters to specific values} 
         &\multirow{2}{1.5cm}{\centering - }&\multirow{2}{1.5cm}{\centering +/-} & \multirow{2}{1.5cm}{\centering + }& \multirow{2}{1.5cm}{\centering + }
        & \multirow{2}{1.5cm}{\centering + }&\multirow{2}{1.5cm}{\centering +/- } &\multirow{2}{1.5cm}{\centering +/- } & \multirow{2}{1.5cm}{\centering + }\\ \\
        \multirow{2}{8cm}{\hphantom{000}Constrain model parameters}  &\multirow{2}{1.5cm}{\centering - }&\multirow{2}{1.5cm}{\centering +/- } & \multirow{2}{1.5cm}{\centering \ding{109} } & \multirow{2}{1.5cm}{\centering \ding{109} } &
        \multirow{2}{1.5cm}{\centering + }&\multirow{2}{1.5cm}{\centering - }&\multirow{2}{1.5cm}{\centering - } & \multirow{2}{1.5cm}{\centering + }\\ \\
         \multirow{2}{8cm}{\hphantom{000}Specify cross-loadings} &\multirow{2}{1.5cm}{\centering - }&\multirow{2}{1.5cm}{\centering - }&\multirow{2}{1.5cm}{\centering - }& \multirow{2}{1.5cm}{\centering - }
         & \multirow{2}{1.5cm}{\centering + }& \multirow{2}{1.5cm}{\centering + }& \multirow{2}{1.5cm}{\centering + }
         & \multirow{2}{1.5cm}{\centering + }\\ \\
        \multirow{2}{8cm}{\hphantom{000}Specify cross-weights} 
        &\multirow{2}{1.5cm}{\centering - }&\multirow{2}{1.5cm}{\centering - }&\multirow{2}{1.5cm}{\centering - }& \multirow{2}{1.5cm}{\centering - }& \multirow{2}{1.5cm}{\centering \ding{109} } & \multirow{2}{1.5cm}{\centering - } & \multirow{2}{1.5cm}{\centering - } &\multirow{2}{1.5cm}{\centering + }\\ \\
        \multirow{2}{8cm}{\hphantom{000}Specify covariances between measurement errors} &\multirow{2}{1.5cm}{\centering +/- } & \multirow{2}{1.5cm}{\centering + }&\multirow{2}{1.5cm}{\centering  + }& \multirow{2}{1.5cm}{\centering + }& \multirow{2}{1.5cm}{\centering + }& \multirow{2}{1.5cm}{\centering + }& \multirow{2}{1.5cm}{\centering + }& \multirow{2}{1.5cm}{\centering + }\\ \\
        \multirow{2}{8cm}{\hphantom{000}Specify covariances between composite indicators and \hphantom{000}measurement errors} &\multirow{2}{1.5cm}{\centering - } &\multirow{2}{1.5cm}{\centering - }&\multirow{2}{1.5cm}{\centering - }& \multirow{2}{1.5cm}{\centering - }& \multirow{2}{1.5cm}{\centering \ding{109} } & \multirow{2}{1.5cm}{\centering - } &\multirow{2}{1.5cm}{\centering - }
        & \multirow{2}{1.5cm}{\centering + }\\ \\
        \multirow{2}{8cm}{\hphantom{000}Specify composites as dependent variables in the \hphantom{000}structural model}& \multirow{2}{1.5cm}{\centering + }& \multirow{2}{1.5cm}{\centering + }& \multirow{2}{1.5cm}{\centering + }& \multirow{2}{1.5cm}{\centering + }& \multirow{2}{1.5cm}{\centering - }& \multirow{2}{1.5cm}{\centering + }&\multirow{2}{1.5cm}{\centering + } & \multirow{2}{1.5cm}{\centering + }\\ \\
        \multirow{2}{8cm}{\hphantom{000}Specify covariances between composites and other \hphantom{000}variables in the model} & \multirow{2}{1.5cm}{\centering + }& \multirow{2}{1.5cm}{\centering + }& \multirow{2}{1.5cm}{\centering + }& \multirow{2}{1.5cm}{\centering + }& \multirow{2}{1.5cm}{\centering - }& \multirow{2}{1.5cm}{\centering + }&\multirow{2}{1.5cm}{\centering + }& \multirow{2}{1.5cm}{\centering + }\\ \\
        \multirow{2}{8cm}{\hphantom{000}Specify the model in terms of the parameters of interest} & \multirow{2}{1.5cm}{\centering + }& \multirow{2}{1.5cm}{\centering + }& \multirow{2}{1.5cm}{\centering + }& \multirow{2}{1.5cm}{\centering + }& \multirow{2}{1.5cm}{\centering + }& \multirow{2}{1.5cm}{\centering + }&\multirow{2}{1.5cm}{\centering - }& \multirow{2}{1.5cm}{\centering + }\\ \\
        \multirow{2}{8cm}{\textit{Model Estimation}}\\ \\
        \multirow{2}{8cm}{\hphantom{000}Estimate the composite weights freely} & \multirow{2}{1.5cm}{\centering + }& \multirow{2}{1.5cm}{\centering + }& \multirow{2}{1.5cm}{\centering + }& \multirow{2}{1.5cm}{\centering + }& \multirow{2}{1.5cm}{\centering + }&\multirow{2}{1.5cm}{\centering - }& \multirow{2}{1.5cm}{\centering + }& \multirow{2}{1.5cm}{\centering + }\\ \\
        \multirow{2}{8cm}{\hphantom{000}Estimate unstandardized path coefficients} & \multirow{2}{1.5cm}{\centering - }&\multirow{2}{1.5cm}{\centering + }& \multirow{2}{1.5cm}{\centering - }&\multirow{2}{1.5cm}{\centering - }&\multirow{2}{1.5cm}{\centering + }& \multirow{2}{1.5cm}{\centering + }& \multirow{2}{1.5cm}{\centering + }& \multirow{2}{1.5cm}{\centering + }\\ \\
               \multirow{2}{8cm}{\textit{Model Assessment}}\\ \\
         \multirow{2}{8cm}{\hphantom{000}Calculate analytic asymptotic standard errors}& \multirow{2}{1.5cm}{\centering \ding{109} } & \multirow{2}{1.5cm}{\centering \ding{109} } & \multirow{2}{1.5cm}{\centering \ding{109} } & \multirow{2}{1.5cm}{\centering + }& \multirow{2}{1.5cm}{\centering + }& \multirow{2}{1.5cm}{\centering + }& \multirow{2}{1.5cm}{\centering + } & \multirow{2}{1.5cm}{\centering + }\\ \\
        \multirow{2}{8cm}{\hphantom{000}Draw direct inference about the model parameters of \hphantom{000}interest} & \multirow{2}{1.5cm}{\centering + }& \multirow{2}{1.5cm}{\centering + }& \multirow{2}{1.5cm}{\centering + }& \multirow{2}{1.5cm}{\centering + }& \multirow{2}{1.5cm}{\centering + }& \multirow{2}{1.5cm}{\centering + }&\multirow{2}{1.5cm}{\centering  - }& \multirow{2}{1.5cm}{\centering + }\\ \\
        \bottomrule
    \end{tabular}
    \end{spacing}
     \begin{tablenotes}
      \small
      \item $+$: The respective approach has the capability; $-$: The respective approach does not have the capability; $+/-$ In some situations the respective approach has the capability, in some situations it does not; \ding{109}: It is unclear whether the respective approach has the capability.
    \end{tablenotes}
\end{sidewaystable}
  \end{adjustbox}

To demonstrate our approach, we conducted a scenario analysis using a sample accurately representing the population.
This scenario analysis confirmed that our approach to SEM with common factors and composites employing the ML estimator can retrieve the population parameters in such a scenario.
Moreover, a small simulation study confirmed that FC-SEM when using the ML estimator yields parameter estimates that are, on average, close to the population counterparts in finite samples.
Additionally, we presented an empirical example to illustrate our method and demonstrate how it can be applied using the open-source R package lavaan.
We further compared the results of our approach with those of other methods for estimating structural equation models involving composites and common factors.
This comparison demonstrates that the standardized parameter estimates are similar among the approaches.

We have demonstrated that our approach allows for reliance on approaches developed in the traditional SEM framework in terms of model estimation and model assessment.
Subsequently, conclusions about these approaches, e.g., in terms of performance in finite samples, can be transferred to our approach.
Furthermore, our model specification allows us to apply developments of the traditional SEM framework such as handling missing data, multi-group analysis, and examining mediation effects.
Future research should explore the performance of these techniques for our approach in more detail.
Generally, assessing the overall model fit is well explored for structural equation models involving only common factors.
However, some of the criteria are not well suited to assess models involving composites \parencite{Henseler2025, Schuberth2020}.
Specifically, measures for assessing validity or reliability are designed for common factors, limiting their applicability for composite models. 
Currently, research on the use of model fit criteria for composite models is rather limited \parencite{Hwang2023,Schuberth2023c}.
Future research should elaborate more on the applicability and interpretability of different model assessment criteria for structural equation models with common factors and composites. 
Additionally, new criteria should be developed to assess such models.
Moreover, currently indicators are restricted to be either composite or common factor indicators. 
Future research should elaborate on including indicators as measures for common factors and indicators of composites simultaneously.
Further, currently our approach is only implemented in the open-source software lavaan which currently limits its availability to R users.
Future research hopefully integrates it in other software packages.
Finally, it is crucial to investigate the assumption that composite indicators are free from measurement error.
Given the likelihood of measurement errors in empirical research, strategies to incorporate and address such errors within our framework should be developed.

In conclusion, our approach significantly broadens the applicability of SEM for studying structural equation models involving composites and common factors, making it more accessible and practical for a wider research audience.

\noindent\textbf{Open Practices Statement}:
\noindent
The R code used for the scenario analysis, for the simulation and for the empirical example are available via:
\url{https://osf.io/9tm2y/?view_only=59d9792bab994892a735e4efc763b511}

\printbibliography

\section*{Appendix}

\subsection*{Model Specification}
To obtain the composite loadings, regressions of the composites on their related indicators need to be performed.
For a composite $\eta_i$ formed by a set of indicators $\bm y_i$, it holds that:
\begin{align}
    \eta_i &= \bm w_i' \bm y_i\\
    \bm y_i &= \bm \lambda_i \eta_i \\
    \Rightarrow \bm \lambda_i &= \frac{\text{cov}(\bm y_i, \eta_i)}{\text{var}(\eta_i)} \\
    &= \frac{\text{cov}(\bm y_i, \bm w_i' \bm y_i) }{\bm w_i' \text{\textbf{V}}(\bm y_i)\bm w_i} \\
    &= \frac{\text{\textbf{V}}(\bm y_i)\bm w_i}{\bm w_i' \text{\textbf{V}}(\bm y_i)\bm w_i}
\end{align}
Combining this for all composites yields to:
\begin{align*}
    \bm \Lambda^c &= \bm T \bm W (\bm W' \bm T \bm W)^{-1}
\end{align*}
\clearpage
\subsection*{Scenario Analysis}
The matrix $\bm T$ contains the unrestricted variances and covariances of the composite indicators.
All remaining entries are set to 0, i.e.:
\begin{align}
    \bm T &= \begin{pmatrix}
        6 &  2& 1.2 & 0 & 0 & 0 & 0 \\
        2& 5& 1.5 & 0 & 0 & 0 & 0\\
        1.2& 1.5& 2& 0 & 0 & 0 & 0\\
        0 & 0 & 0 & 7& 0.9& 0.5& 1.2\\ 
        0 & 0 & 0 &0.9& 3& 0.7& 1.8\\
        0 & 0 & 0 &0.5& 0.7& 2& 0.5\\
        0 & 0 & 0 &1.2& 1.8& 0.5& 5
    \end{pmatrix}
\end{align}
The composite matrix $\bm W$ contains the composite weights in its columns, i.e.:
\begin{align*}
\bm W =
    \begin{pmatrix}
        1.0 & 0.0\\
        0.4 & 0.0 \\
        0.6 & 0.0 \\
        0.0 & 1.0 \\
        0.0 & 0.2 \\
        0.0 & 0.5 \\
        0.0 & 0.3
    \end{pmatrix}
\end{align*}
\clearpage
Thus, the composite loading matrix is given as:
\begin{align}
    \bm \Lambda^c = \bm T \bm W(\bm W' \bm T \bm W)^{-1} = \begin{pmatrix}
        0.67 & 0.00 \\
        0.43 & 0.00 \\
        0.27 & 0.00\\
        0.00 & 0.77 \\
        0.00 & 0.24 \\
        0.00 & 0.18 \\
        0.00 & 0.33
    \end{pmatrix}
\end{align}

The relations between the constructs and the indicators are therefore given as:
\begin{align}
    \begin{pmatrix}
        \eta_1 \\
        \eta_2 \\
        \eta_3 \\
        \eta_4
    \end{pmatrix} &= \begin{pmatrix}
        1.00 & 0.00 & 0.00 & 0.00\\
        0.80 & 0.00 & 0.00 & 0.00\\
        0.90 & 0.00 & 0.00 & 0.00\\
        0.00 & 1.00 & 0.00 & 0.00\\
        0.00 & 0.70 & 0.00 & 0.00 \\
        0.00 & 0.80 & 0.00 & 0.00 \\
        0.00 & 0.00 & 0.67 & 0.00 \\
        0.00 & 0.00 &  0.43 & 0.00 \\
        0.00 & 0.00 & 0.27 & 0.00\\
        0.00 & 0.00 & 0.00 & 0.77 \\
        0.00 & 0.00 & 0.00 & 0.24 \\
        0.00 & 0.00 & 0.00 & 0.18 \\
        0.00 & 0.00 & 0.00 & 0.33
    \end{pmatrix} \begin{pmatrix}
        y_1 \\
        y_2 \\
        y_3 \\
        y_4 \\
        y_5\\
        y_6 \\
        y_7 \\
        y_8 \\
        y_9 \\
        y_{10}\\
        y_{11}\\
        y_{12}\\
        y_{13}
    \end{pmatrix} + \begin{pmatrix}
        \varepsilon_1\\
        \varepsilon_2 \\
        \varepsilon_3\\
        \varepsilon_4\\
        \varepsilon_5 \\
        \varepsilon_6 \\
        0 \\
        0\\
        0\\
        0\\
        0\\
        0\\
        0
    \end{pmatrix}
\end{align}

To obtain the variances of the error terms related to composites, we need to take into account that the variances of the composites are given by the composite model, i.e.:
\begin{align*}
      \eta_3 &= 1 \cdot y_7 + 0.4\cdot  y_8 + 0.6 \cdot  y_9 = (1.0\ 0.4 \ 0.6)  (y_7 \ y_8\ y_9)' \\
      \Rightarrow \quad \text{var}(\eta_3) &= (1.0\ 0.4 \ 0.6) \text{V}(y_7, y_8, y_9) (1.0\ 0.4 \ 0.6) ' = 11.28\\
    \eta_4 &= 1 \cdot y_{10} + 0.2 \cdot y_{11} + 0.5 \cdot y_{12} + 0.3 \cdot y_{13} = (1.0\ 0.2\ 0.5\ 0.3) (y_{10}\ y_{11}\ y_{12} \ y_{13})'\\
    \Rightarrow \quad \text{var}(\eta_4) &= (1.0\ 0.2\ 0.5\ 0.3) \text{V}(y_{10}, y_{11}, y_{12},y_{13}) (1.0\ 0.2\ 0.5\ 0.3)' = 10.156
\end{align*}
As $\eta_4$ is also a dependent variable in the structural model, it holds:
\begin{align*}
    \eta_4 &= 0.6 \cdot \eta_1 + + 0.5\cdot \eta_3 + \zeta_4 \\
    \Rightarrow \quad \text{var}(\eta_4) &= 0.6^2 \cdot \text{var} (\eta_1) + 0.5^2\cdot  \text{var}(\eta_3) + 2 \cdot 0.5 \cdot 0.6\cdot  \text{cov}(\eta_1, \eta_3) + \text{var}(\zeta_4) \\    
\end{align*}
Thus, the variance of $\zeta_4$ is given as:
\begin{align*}
    \text{var}(\zeta_4) &=\text{var}(\eta_4)  - (0.6^2\cdot  \text{var} (\eta_1) + 0.5^2\cdot  \text{var}(\eta_3) + 2 \cdot 0.5 \cdot 0.6\cdot  \text{cov}(\eta_1, \eta_3))= 6.56
\end{align*}
Therefore, the model-implied variance-covariance matrix of the structural error terms is given as:
\begin{align}
    \text{var}(\bm \zeta) = \begin{pmatrix}
        2.00 & 0.00 & 0.10 & 0.00\\
        0.00 & 1.50 & 0.00 & 0.00\\
        0.10 & 0.00 & 11.28 & 0.00\\
        0.00 & 0.00 & 0.00 & 6.56
    \end{pmatrix}
\end{align}
Consequently, the model-implied variance-covariance matrix of the constructs is given as:
\begin{align*}
    \text{var}(\bm \eta)& = (\bm I - \bm B)^{-1} \text{var}(\bm \zeta)(\bm I - \bm B)'^{-1} \\&= \begin{pmatrix}
        1.0 & 0.0 & 0.0 & 0.0\\
        -0.2 & 1.0 & -0.1 & -0.3 \\
        0.0 & 0.0 & 1.0 & 0.0\\
        -0.6 & 0.0 & -0.5 & 1.0
    \end{pmatrix}^{-1} \begin{pmatrix}
        2.00 & 0.00 & 0.10 & 0.00\\
        0.00 & 1.50 & 0.00 & 0.00\\
        0.10 & 0.00 & 11.28 & 0.00\\
        0.00 & 0.00 & 0.00 & 6.56
    \end{pmatrix} 
    \begin{pmatrix}
        1.0 & -0.2 & 0.0 & -0.6\\
        0.0 & 1.0 & 0.0 & 0.0 \\
        0.0 & -0.1 & 1.0 & -0.5\\
        0.0 & -0.3 & 0.0 & 1.0
    \end{pmatrix}^{-1} \\&= \begin{pmatrix}
        2.00 &0.78&  0.10 & 1.25\\
0.78& 3.10 & 2.86 & 3.87\\
0.10& 2.86 &11.28 & 5.70\\
1.25& 3.87 & 5.70 &10.16
    \end{pmatrix}
    \end{align*}

While the variance-covariance matrix of the measurement errors of the common factor indicators $\bm \Theta^f$ is given as a diagonal matrix with 0.5 on the diagonal, the matrix $\bm \Theta^c$ can be obtained as follows:
\begin{align*}
    \bm \Theta^c = \bm T - \bm \Lambda^c \bm W' \bm T \bm W \bm \Lambda^{c'} = \begin{pmatrix}
0.99 &-1.27 &-0.8 & 0.00 & 0.00  &0.00 & 0.00\\
 -1.27 & 2.87 & 0.2  &0.00 & 0.00  &0.00 & 0.00\\
 -0.80 & 0.20 & 1.2 & 0.00  &0.00  &0.00 & 0.00\\
 0.00  &0.00 & 0.00&  1.02 &-0.93 &-0.87 &-1.34\\
 0.00  &0.00 & 0.00 &-0.93 & 2.44 & 0.28 & 1.02\\
 0.00  &0.00 & 0.00 &-0.87 & 0.28  &1.68 &-0.08\\
 0.00  &0.00 & 0.00& -1.34 & 1.02 &-0.08 & 3.92\\
    \end{pmatrix}
\end{align*}
\clearpage
Finally, the model-implied variance-covariance matrix of the indicators is given as:
\begin{align*}
    \bm \Sigma(\bm y) &= \begin{pmatrix}
2.50 \\
1.60& 1.78 \\
1.80 &1.44 &2.12 \\
0.78 &0.63 &0.71 &3.60\\
0.55 &0.44 &0.49 &2.17 &2.02\\
0.63 &0.50 &0.57 &2.48 &1.74 &2.49\\
0.07 &0.05 &0.06 &1.91 &1.33 &1.52 &6.00\\
0.04 &0.03 &0.04 &1.24 &0.87 &0.99 &2.00 &5.00\\
0.03 &0.02 &0.02 &0.76 &0.53 &0.61 &1.20 &1.50 &2.00\\
0.96 &0.77 &0.86 &2.97 &2.08 &2.37 &2.91 &1.90 &1.16 &7.00\\
0.29 &0.24 &0.26 &0.91 &0.64 &0.73 &0.89 &0.58 &0.36 &0.90 &3.00\\
0.22 &0.18 &0.20 &0.68 &0.48 &0.55 &0.67 &0.44 &0.27 &0.50 &0.70 &2.00 \\
0.41 &0.33 &0.37 &1.26 &0.88 &1.01 &1.24 &0.81 &0.49 &1.20 &1.80 &0.50 &5.00
    \end{pmatrix}
\end{align*}

\end{document}